\documentclass{article}

\usepackage{amssymb, amsmath}

\begin{document}

\title{Flavour-mixing gauge field theory of massive Majorana neutrinos\newline}

\author{Eckart Marsch}

\date{}
\maketitle

\noindent
{\bf Institute for Experimental and Applied Physics, Christian
Albrechts University at Kiel, Leibnizstr. 11, 24118 Kiel, Germany
\newline\newline\newline}

\noindent
{\bf Abstract \newline} A gauge-field theory for massive neutral
particles is developed on the basis of the real four-component Majorana
equation. By use of its spin operator, a purely imaginary representation of
the SU(2) algebra can be defined, which gives a covariant derivative that is
real. Such a coupling to the gauge field preserves the real nature of the
Majorana equation even when including interactions. As the associated isospin
is four-dimensional, this procedure introduces four intrinsic degrees of
freedom to the Majorana field, which may be related to four flavours. The
main aim is to describe here the mathematical possibility for coupling
Majorana particles with a gauge field which resembles that of the weak
interaction. By adding a fourth member to the family, flavour could become a
dynamic trait of the neutral Majorana particles, and thus lead to a dynamic
understanding of mixing.

\newpage

\section{Introduction}
\label{s1}

According to the canonical standard model of elementary particle physics, the
leptons and quarks come in three flavours, are massless and thus obey chiral
symmetry, but then they can acquire mass through the Higgs mechanism (see,
e.g., the text books \cite{lee, kaku, das}). The Dirac equation \cite{dirac}
is fundamental in all this and well understood, however the nature of the
neutrinos involved remains less clear. Are they Dirac fermions or Majorana
\cite{majorana} particles? The physics of neutrinos and its application to
particle astrophysics remains a very active research area. In the past
neutrinos were often described by the massless Weyl \cite{weyl} equations
involving only two-component Pauli \cite{pauli} spinors. However, since
convincing empirical evidence \cite{fukugita} for the finite neutrino masses
and their associated oscillations \cite{mohapatra, kayser1} had been found in
the past decades, massive neutrinos have been discussed, and furthermore
another very massive neutrino species or sterile ones have been considered in
four-neutrino models \cite{kalliomaki}, for example to explain the light
neutrinos masses by the see-saw mechanism \cite{mohapatra}. The tensions with
the three-neutrino paradigm have recently been discussed by Kayser
\cite{kayser2}, who also reviewed the key arguments in support for neutrinos
being Majorana fermions \cite{kayser3}. Therefore the Majorana equation with
various mass terms has found strong attention, either in its complex
two-component (see the recent review by Dreiner et al. \cite{dreiner}) or
real four-component form, and been used in modern quantum field theory for
the description of massive neutrinos. Thus theoretical reasoning and new
empirical results from laboratory as well as cosmology suggest the possible
existence of a fourth neutrino flavour. The state of affairs (as of 2006) and
the research perspectives are described in the comprehensive review by
Mohapatra and Smirnov \cite{smirnow}.

The purpose of the present paper is to show that for massive neutral Majorana
particles (perhaps representing the observed neutrinos) a gauge-field theory
based on the real four-component Majorana equation is feasible. Using its
spin operator, which obeys the angular momentum algebra like SU(2) but is
purely imaginary, we can define an appropriate symmetry group which gives a
connection that is real, and thus the coupling to the gauge field (also real)
does keep the real nature of the Majorana equation even when including
interactions. As the isospin is four-dimensional, this procedure introduces
four intrinsic degrees of freedom to the Majorana field, the interpretation
of which remains open to speculation. Here the main aim is to describe the
mathematical possibility for coupling the Majorana particles with a gauge
field, which resembles that of the weak interactions between leptons and
quarks \cite{weinberg, glashow, salam}. Although the neutrinos are presently
known to come in three flavours, we will here argue that, by adding a fourth
member to the family, flavour could become a dynamic trait of the neutral
Majorana particles and be related to the gauge theory described subsequently.

The paper is organized as follows. We discuss the relevant key aspects of the
Majorana equation (its eigenfunctions are given in the appendix) in the next
section, which is more of a tutorial nature. Then the real Lorentz
transformation is given, in which the purely imaginary Majorana spin operator
occurs prominently. In our opinion, it strongly suggests itself as the
adequate isospin for the SU(2)-like algebra adopted in the gauge symmetry
considered in the subsequent section. When assuming four flavours, the
resulting gauge theory provides a dynamic picture of flavour mixing, i.e.
quantum flavour dynamics, and likely also new insights into neutrino
oscillations.

\section{The Majorana equation}
\label{s2}

In this introductory section, we consider the Dirac equation in its Majorana
representation. We use standard symbols, notations and definitions, and
conventionally units of $\hbar=c=1$, with the four-momentum operator denoted
as $P_\mu=(E, -\mathbf{p}) =\mathrm{i}\partial_\mu =
\mathrm{i}(\partial/\partial t,\partial/\partial \mathbf{x})$, which acts on
the spinor wave function $\psi(\mathbf{x},t)$. The particle mass is $m$. The
four-vector $\gamma^\mu$ consists of the four Dirac gamma matrices that come
in various representations. As first found out by Majorana \cite{majorana},
there exists a purely imaginary representation such that $\gamma^\mu =
\mathrm{i} \bar{\gamma}^\mu$, which makes the Dirac equation real. We simply
refer to it as Majorana equation in the remainder of this work. This real
equation reads
\begin{equation}
\label{eq:1} \bar{\gamma}^\mu \partial_\mu\psi +  m\psi = 0.
\end{equation}

The Majorana equation involves the below defined $4\times4$ matrices.
Throughout the text we will use top-barred symbols to indicate real
$4\times4$ matrices. The gamma matrices can after \cite{marsch1} be defined
as
\begin{equation}
\label{eq:2}
\bar{\gamma}^\mu=
\left(
\left(
\begin{array}{cc}
0 & \gamma \\
\gamma & 0  \\
\end{array}
\right),
\left(
\begin{array}{cc}
0 & -\alpha  \\
-\alpha &  0 \\
\end{array}
\right),
\left(
\begin{array}{cc}
1  &  0  \\
0  &  -1  \\
\end{array}
\right),
\left(
\begin{array}{cc}
0  & \beta  \\
\beta &  0  \\
\end{array}
\right)
\right).
\end{equation}
The gamma matrices mutually anticommute and obey:
$\bar{\gamma}^\mu\bar{\gamma}^\nu + \bar{\gamma}^\nu\bar{\gamma}^\mu = -
2g^{\mu\nu}$, with euclidian metric $g^{\mu\nu}$. The three associated real
$2\times2$ matrices read
\begin{equation}
\label{eq:3} \alpha = \left(
\begin{array}{cc}
1  &  0  \\
0  & -1  \\
\end{array}
\right), \;\;\; \beta= \left(
\begin{array}{cc}
0  &  1 \\
1  &  0 \\
\end{array}
\right), \;\;\; \gamma= \left(
\begin{array}{cc}
0  &  -1  \\
1  &   0  \\
\end{array}
\right).
\end{equation}
Note that $\gamma = \beta \alpha$, and that all three matrices mutually
anti-commute with each other, just like the Pauli matrices do. All what we
need in the following are the algebraic properties of these $2\times2$
matrices involved, like $\alpha\beta+\beta\alpha=0$,
$\alpha\gamma+\gamma\alpha=0$, and $\beta\gamma+\gamma\beta=0$, and that
$\alpha^2=\beta^2=1$ and $\gamma^2=-1$. The three Pauli matrices have their
standard form, and are given by $\sigma_{\rm x}=\beta$, $\sigma_{\rm y}=
\mathrm{i}\gamma$,and $\sigma_{\rm z}=\alpha$.

The four-component Majorana equation (\ref{eq:1}) is a direct consequence of,
and fully equivalent to, the two-component complex Majorana equation. It
involves only the Pauli matrix operators acting on a two-component spinor
$\phi$, but introduces subtle complications that are caused by the spin-flip
operator $\tau$, and reads as follows:
\begin{equation}
\label{eq:5} \mathrm{i}\left( \frac{\partial}{\partial t} +
\boldsymbol{\sigma} \cdot \frac{\partial}{\partial \mathbf{x}} \right)
\phi(\mathbf{x},t) = m \tau \phi(\mathbf{x},t).
\end{equation}
It can be derived without invoking the Dirac equation in the first place, as
was demonstrated by Marsch \cite{marsch1,marsch2} recently. As shown years
ago by Case \cite{case}, this equation can be also derived from Dirac's
equation in its chiral form.

Above in (\ref{eq:5}) we defined an important operator that is not of pure
algebraic nature but involves the complex-conjugation operator named as
$\mathbb{C}$. This antihermitian operator called $\tau$ is defined as $\tau =
\sigma_y \mathbb{C}$, and obeys $\tau^{-1}= -\tau$. The operation of $\tau$
on the spin vector $\boldsymbol{\sigma}$ leads to its inversion, i.e., the
operation $\tau \boldsymbol{\sigma} \tau^{-1} = -\boldsymbol{\sigma}$ yields
a spin flip. We also have $\tau \mathrm{i}=-\mathrm{i}\tau$ because of the
action of $\mathbb{C}$. Therefore, $\tau$ anti-commutes with the momentum
four-vector operator $P^\mu$. Moreover, a simple phase factor like
$\exp(\mathrm{i}\theta)$ (with some angle $\theta$) does not commute with the
mass term on the right-hand side of (\ref{eq:5}). Consequently, that equation
can neither describe electromagnetic interactions nor most other complex
gauge-field couplings in which the imaginary unit $\mathrm{i}$ appears
explicitly.

To avoid the complications introduced by the operator $\tau$, we prefer to
work with the real four-dimensional Majorana equation (\ref{eq:1}), which we
quote again but now in conventional Hamiltonian form as used by Majorana [5]
himself as follows
\begin{equation}
\label{eq:6} \left( \frac{\partial}{\partial t} + \bar{
\boldsymbol{\alpha}} \cdot \frac{\partial}{\partial \mathbf{x}} \right)
\psi(\mathbf{x},t) = m \bar{\beta}\, \psi(\mathbf{x},t),
\end{equation}
which formally resembles the complex two-component version (\ref{eq:5}). This
last equation is as usually obtained by multiplying the manifestly covariant
form (\ref{eq:1}) from the left by $\bar{\gamma}^0=\bar{\beta}$ and defining
$\bar{\alpha}^\mu=-\bar{\gamma}^0\bar{\gamma}^\mu=(1,\bar{\boldsymbol{\alpha}})$.
The real matrix three-vector $\bar{\boldsymbol{\alpha}}$ is then given as
\begin{equation}
\label{eq:7}
\bar{\boldsymbol{\alpha}} =
\left(
\left(
\begin{array}{cc}
\beta & 0 \\
0 &  \beta \\
\end{array}
\right),
\left(
\begin{array}{cc}
0 & \gamma  \\
-\gamma &  0  \\
\end{array}
\right),
\left(
\begin{array}{cc}
\alpha  & 0  \\
0 &  \alpha \\
\end{array}
\right)
\right),
\end{equation}
and is symmetric, which means equal to its transposed matrix,
$\bar{\boldsymbol{\alpha}} =\bar{\boldsymbol{\alpha}}^\mathrm{T}$. The
eigenfunctions of the Majorana equations (\ref{eq:5}) and (\ref{eq:6}) are
briefly discussed in the appendix Section~\ref{s6}, where it is shown that
the free real Majorana can be decomposed into left- and right-helical
particle and antiparticle components. It is chirally irreducible, though,
unless one considers complex solutions.

At this point, we quote the Lagrangian density of the Majorana field, which
is obtained by inserting the Majorana gamma matrices into the general Dirac
Lagrange density. Note that the factor $\mathrm{i}$ is important here to
ensure that we are dealing with hermitian operators having real eigenvalues.
Yet the factor $\mathrm{i}$ is irrelevant, and does not appear, in the
Majorara equation (\ref{eq:6}) itself. The final result is (where the
superscript T denotes the transposed spinor) given by
\begin{equation}
\label{eq:9}
\mathcal{L}_\mathrm{M} = \mathrm{i} \psi^\mathrm{T} (\bar{\alpha}^\mu \partial_\mu - \bar{\beta} m )\psi.
\end{equation}
Its variation with respect to $\psi^\mathrm{T}$ yields the above equation of
motion (\ref{eq:6}) for the Majorana field. It should here be stressed again
that the Majorana equation has two degrees of freedom less than the complex
Dirac equation in its standard form and describes particle and antiparticles
of opposite mean helicity. So using the real Dirac, i.e. the Majorana,
equation implies and automatically ensures this reduction of the degrees of
freedom (see the appendices Section~\ref{s7} and Section~\ref{s8}).
Throughout the remainder of our paper we will stay with the real description.

\section{Symmetries and Lorentz transformation in Majorana representation}
\label{s3}

Now let us briefly discuss the symmetries of the Majorana equation
(\ref{eq:6}). To prepare this we recall that there is an important symmetry
operator, namely the chiral matrix operator $\gamma^5$, which (see for
example \cite{kaku}) is defined as $\gamma^5=
\mathrm{i}\gamma^0\gamma^1\gamma^2\gamma^3= \mathrm{i} \bar{\gamma}^5$.
Inserting the gamma matrices of equation (\ref{eq:2}) we obtain,
\begin{equation}
\label{eq:8}
\bar{\gamma}^5 =  \bar{\alpha}_\mathrm{x} \bar{\alpha}_\mathrm{y} \bar{\alpha}_\mathrm{z}
=\left(
\begin{array}{cc}
0 & 1 \\
-1 & 0 \\
\end{array}
\right).
\end{equation}
Like the other gamma matrices, $\gamma^5$ is purely imaginary. As a result
one can not, by standard projection, obtain real right- or left-chiral
components of the real eigen-spinor solutions of the Majorana equation
(\ref{eq:6}) as given by (\ref{eq:102}) and (\ref{eq:103}) in the appendix
Section~\ref{s7}.

Obviously, the chirality operator $\bar{\gamma}^5$ by its definition
(\ref{eq:8}) commutes with $\bar{\boldsymbol{\alpha}}$. But we require an
operator like the $\tau$ of Section~\ref{s2} which anticommutes with the
alphas. The product
$\bar{\beta}\bar{\gamma}^5=\bar{\delta}=\bar{\beta}\bar{\alpha}_\mathrm{x}
\bar{\alpha}_\mathrm{y} \bar{\alpha}_\mathrm{z}$ has that desired property
and thus corresponds to $\tau$. So we define:
\begin{equation}
\label{eq:404}
\bar{\delta}
=\left(
\begin{array}{cc}
-\gamma & 0 \\
0 & \gamma \\
\end{array}
\right),
\end{equation}
the square of which is -1 (like the square of $\tau$), and which anticommutes
with $\bar{\beta}$ as well.

Concerning the symmetries of the Majorana equation, we consider in particular
chirality conjugation $\mathcal{C}$, parity $\mathcal{P}$, and time reversal
$\mathcal{T}$ operations. Generally speaking the Majorana equation is
invariant under the symmetry operation $\mathcal{O}$, if the spinor
\begin{equation}
\label{eq:400} \psi^\mathcal{O} = \mathcal{O}\psi
\end{equation}
also fulfils that equation. Therefore, when applying the operation
$\mathcal{O}$ from the left and its inverse $\mathcal{O}^{-1}$ from the
right, whereby the unit operator is given by the decomposition
$\mathcal{O}\mathcal{O}^{-1}=1$, we obtain the result
\begin{equation}
\label{eq:401} \left( \mathcal{O}\frac{\partial}{\partial
t}\mathcal{O}^{-1} + \mathcal{O}(\bar{\boldsymbol{\alpha}} \cdot
\frac{\partial}{\partial \mathbf{x}})\mathcal{O}^{-1}
- m \mathcal{O} \bar{\beta} \mathcal{O}^{-1} \right) \mathcal{O}\psi(\mathbf{x},t)=0.
\end{equation}
We conventionally define the time and space coordinate inversion operations
$\mathbb{T}$ and $\mathbb{P}$ on a spinor $\psi$ by
\begin{equation}
\label{eq:402} \mathbb{T} \psi(\mathbf{x},t)= \psi(\mathbf{x},-t),
\end{equation}
\begin{equation}
\label{eq:403} \mathbb{P} \psi(\mathbf{x},t)= \psi(-\mathbf{x},t),
\end{equation}
and also recall the complex conjugation operation $\mathbb{C}$, which gives
$\mathbb{C} \mathrm{i} \mathbb{C}^{-1} = - \mathrm{i}$, and but which here
has no role to play as everything is real. With these preparations in mind,
it is easy to see which operators provide the requested symmetry operations.
We compose them in the Table~\ref{table:1}. To complete the operator algebra,
it is important to note the following commutation relations. Of course, the
coordinate reversal operators $\mathbb{T}$ and $\mathbb{P}$ commute with
$\bar{\alpha}$, $\bar{\beta}$ and $\bar{\delta}$.

\begin{table}
\caption{Symmetry operations} \vspace{0.5 em}
\begin{tabular}{lccc}
\hline
Operation      &  Time reversal      &      Parity   &  Chirality conjugation  \\
Operator & $\bar{\delta}\mathbb{T}$  & $ \bar{\beta} \mathbb{P}$ &   $\bar{\delta}$  \\
\hline
\end{tabular}
\label{table:1}
\end{table}

Let us first consider in (\ref{eq:401}) the time reversal,
$\mathcal{O}=\mathcal{T}=\bar{\delta}\mathbb{T}$. Apparently, it affects the
mass term via $\bar{\beta}$ as well as $\bar{\boldsymbol{\alpha}}$, where the
signs are reversed, and also changes the sign of the first term. Therefore,
also $\psi^\mathcal{T}=\bar{\delta}\psi(\mathbf{x},-t)$ solves the Majorana
equation (\ref{eq:6}). The parity operation
$\mathcal{O}=\mathcal{P}=\bar{\beta}\mathbb{P}$ commutes with the mass term
and does not affect the first term in (\ref{eq:6}), and it also leaves the
momentum term invariant since $\boldsymbol{\alpha}$ and $\mathbf{x}$ both
change signs together. Therefore, also
$\psi^\mathcal{P}=\bar{\beta}\psi(-\mathbf{x},t)$ solves the Majorana
equation. Finally, we consider chirality conjugation defined as
$\mathcal{O}=\mathcal{C}=\bar{\delta}$. It changes the signs of the alpha and
beta terms in (\ref{eq:6}) with no net effect on the time-derivate term.
Therefore, $\psi^\mathcal{C}=\bar{\delta}\psi(\mathbf{x},t)$ does not solve
the Majorana equation, but in fact its chirality-conjugated version which is
only derived in the appendix Section~\ref{s8}. In conclusion, the symmetry
operations of Table~\ref{table:1} work in a transparent way on the Majorana
equation.

To obtain the spin operator of the Majorana field, we now consider the
Lorentz group generators, which in its four-component spinor representation
are known [5] to be given by the commutator $\sigma^{\mu\nu}
=\frac{\mathrm{i}}{2} [\gamma^\mu,\gamma^\nu]$, which yields in Majorana
representation with the gamma matrices as given in (\ref{eq:2}) the
subsequent matrix tensor:
\begin{equation}
\label{eq:10}
\sigma^{\mu\nu} = \left(
\begin{array}{cccc}
0 & \mathrm{i} \bar{\alpha}_\mathrm{x} & \mathrm{i} \bar{\alpha}_\mathrm{y}&  \mathrm{i} \bar{\alpha}_\mathrm{z}\\
 -\mathrm{i} \bar{\alpha}_\mathrm{x}  & 0          &  \Sigma_\mathrm{z}  &  -\Sigma_\mathrm{y}  \\
 -\mathrm{i} \bar{\alpha}_\mathrm{y}  &  -\Sigma_\mathrm{z} &  0         &  \Sigma_\mathrm{x}   \\
 -\mathrm{i} \bar{\alpha}_\mathrm{z}  &   \Sigma_\mathrm{y} &  -\Sigma_\mathrm{x} &        0    \\
\end{array}
\right).
\end{equation}
Note that these tensor elements are $4\times4$ matrices. Thus we obtained by
definition the spin operator $\boldsymbol{\Sigma}$ of the Majorana field,
which unlike the spin operators in the Dirac or Weyl representation, is not
diagonal and purely imaginary. It reads:
\begin{equation}
\label{eq:11}
\boldsymbol{\Sigma}= \mathrm{i} \bar{\boldsymbol{\Sigma}} = \mathrm{i}
\left(
\left(
\begin{array}{cc}
0  &  -\beta \\
\beta &  0\\
\end{array}
\right),
\left(
\begin{array}{cc}
\gamma   & 0 \\
0  & \gamma  \\
\end{array}
\right),
\left(
\begin{array}{cc}
0  & -\alpha  \\
\alpha &  0  \\
\end{array}
\right)
\right).
\end{equation}

The spin operator does not commute with $\bar{\boldsymbol{\alpha}}$ but with
$\bar{\beta}$. Therefore its eigenfunctions are not eigenfunctions of the
Majorana equation. Remember that $\bar{\boldsymbol{\alpha}}$ in turn
anticommutes with $\bar{\beta}$. Note further that the spin operator is of
course hermitian, and as a result we have
$\bar{\boldsymbol{\Sigma}}^\mathrm{T}=-\bar{\boldsymbol{\Sigma}}$, whereas
$\bar{\boldsymbol{\alpha}}^\mathrm{T}=\bar{\boldsymbol{\alpha}}$. We also
recall that the above spin matrices obey the same relations like the Pauli
matrices, i.e. $[\Sigma_\mathrm{x}, \Sigma_\mathrm{y}]=
2\mathrm{i}\,\Sigma_\mathrm{z}$, where indices can be permuted in a cyclic
way. So the Majorana spin operator can as usually be defined as
$\mathbf{S}_\mathrm{M}= \frac{1}{2}\bar{\boldsymbol{\Sigma}}$, which forms
the SU(2) Lie algebra, also generating the familiar SU(2) Lie group of the
weak interactions. Finally, by using the $\bar{\boldsymbol{\alpha}}$ and
$\bar{\boldsymbol{\Sigma}}$ matrices, the Lorentz transformation operator
$\Lambda(\boldsymbol{\vartheta}, \boldsymbol{\varphi}$) turns out be real,
and the related matrix can then be written as:
\begin{equation}
\label{eq:12}
S(\Lambda)
= \exp \left(\frac{1}{2} \bar{\boldsymbol{\alpha}}\cdot\boldsymbol{\vartheta} +
\frac{1}{2} \bar{\boldsymbol{\Sigma}}\cdot\boldsymbol{\varphi} \right),
\end{equation}
where the vector angle $\boldsymbol{\vartheta}$ refers to genuine Lorentz
transformations and $\boldsymbol{\varphi}$ to proper spatial rotations. Note
that $S(\Lambda)^\mathrm{T}\bar{\beta}S(\Lambda) = \bar{\beta}$, a key
feature which ensures Lorentz invariance of the Lagrangian, which is obvious
for the mass term in particular in the Lagrange density (\ref{eq:9}).

\section{Gauge field theory of flavour dynamics}
\label{s4}

We are now coming to the main theme of this paper, which is the question of
Majorana (neutrino) gauge-field generated dynamics. Let us first consider an
Abelian gauge field $A_\mu(x)$ like in electrodynamics. Conventionally, this
is introduced into the field equation by replacing, according to the minimal
coupling principle, the time-space derivative $\partial_\mu$ by the covariant
derivative
\begin{equation}
\label{eq:13}
D_\mu  = \partial_\mu + \mathrm{i} e A_\mu.
\end{equation}
Concerning, the Majorana equation (\ref{eq:5}), this coupling causes a
serious problem as the corresponding complex phase, like $\exp(\mathrm{i} e
\lambda(x))$, where $\lambda(x)$ is a function of space-time and $e$ the
coupling constant (charge), in the wavefunction does not commute with the
operator $\tau$. Fortunately, there is no such problem with (\ref{eq:6}),
other than the solution spinor now has to be complex, which is not what we
were aiming for to begin with. However, there is a remedy for getting rid of
the unwanted imaginary unit, if we can find a non-Abelian symmetry group
\cite{yang} the matrix representation of which is purely imaginary (and so
electromagnetic interaction is excluded at the outset). The spin algebra as
given by the Majorana spin operator according to (\ref{eq:11}) provides just
the desired representation, which yet is not identical with the fundamental
unitary representation of SU(2). Therefore we will chose (with some coupling
constant $g$) the connection involving the un-normed isospin operator defined
by the matrix three-vector $\mathbf{S}=\mathrm{i}\bar{\mathbf{S}}$ as
follows:
\begin{equation}
\label{eq:14}
D_\mu  = \partial_\mu + \mathrm{i} g \mathbf{S} \cdot \mathbf{A}_\mu
= \partial_\mu - g \bar{\mathbf{S}} \cdot \mathbf{A}_\mu,
\end{equation}
which is real and thus still permits real solutions to be obtained for the
interacting Majorana field and its spinors. The real matrix isospin
three-vector reads explicitly:
\begin{equation}
\label{eq:16}
\bar{\mathbf{S}}=
\left(
\left(
\begin{array}{cccc}
0 & 0 & 0 & -1 \\
0 & 0 & -1 & 0 \\
0 & 1 & 0 & 0 \\
1 & 0 & 0 & 0 \\
\end{array}
\right),
\left(
\begin{array}{cccc}
0 & -1 & 0 & 0 \\
1 & 0 & 0 & 0 \\
0 & 0 & 0 & -1 \\
0 & 0 & 1 & 0 \\
\end{array}
\right),
\left(
\begin{array}{cccc}
0 & 0 & -1 & 0 \\
0 & 0 & 0 & 1 \\
1 & 0 & 0 & 0 \\
0 & -1 & 0 & 0 \\
\end{array}
\right)
\right).
\end{equation}
The spin matrix algebra leads for any real three-vectors $\mathbf{A}$ and
$\mathbf{B}$ to the useful relation:
\begin{equation}
\label{eq:15}
(\bar{\mathbf{S}}\cdot\mathbf{A})(\bar{\mathbf{S}}\cdot\mathbf{B})
=-\mathbf{A}\cdot\mathbf{B}+ \bar{\mathbf{S}}\cdot(\mathbf{A}\times\mathbf{B}).
\end{equation}

As the Majorana matrix representation is four-dimensional, we have thus
introduced in the Majorana field four new internal degrees of freedom related
to the gauge field, which of course requires a physical interpretation. We
return to this issue in the discussion section and here proceed formally.
Note that $\mathbf{S}$ acts on a four-component spinor
\begin{equation}
\label{eq:17}
\Psi=
\left(
\begin{array}{c}
\psi_1 \\
\psi_2 \\
\psi_3 \\
\psi_4 \\
\end{array}
\right).
\end{equation}
Each single component relates to the Majorana equation (\ref{eq:1}) or
(\ref{eq:6}). For the free field the eigenfunction $\psi$ is given in the
appendix Section~\ref{s6}. The extended Majorana Lagrange density, while
including the gauge-field interaction, now has two parts which read:
\begin{equation}
\label{eq:18}
\mathcal{L}_\mathrm{M} + \mathcal{L}_\mathrm{I} = \mathrm{i} \Psi^\mathrm{T}(\bar{\alpha}^\mu (\partial_\mu
- g \bar{\mathbf{S}} \cdot \mathbf{A}_\mu) -\bar{\beta} m )\Psi.
\end{equation}

We can identify in (\ref{eq:18}) a term associated with the particle flux
density. This interaction term can conventionally be written as
\begin{equation}
\label{eq:22}
\mathcal{L}_\mathrm{I}  =  - \mathbf{j}^\mu \cdot \mathbf{A}_\mu,
\end{equation}
where the real isospin current density generated by the SU(2)-like gauge
symmetry is given by
\begin{equation}
\label{eq:23}
\mathbf{j}^\mu = g \mathrm{i} \Psi^\mathrm{T} \bar{\alpha}^\mu \bar{\mathbf{S}} \Psi =
g \Psi^\mathrm{T} \bar{\alpha}^\mu \mathbf{S} \Psi.
\end{equation}

The tensor of the gauge field strength is according to standard theory
\cite{yang, lee} given by the commutator of the connection (\ref{eq:14}),
i.e. we have
\begin{equation}
\label{eq:19}
F_{\mu\nu} = \frac{\mathrm{i}}{g}[D_\mu,D_\nu]
= \mathrm{i}\bar{\mathbf{S}}\cdot\mathbf{F}_{\mu\nu},
\end{equation}
where use has been made of (\ref{eq:15}). Thus the antisymmetric field tensor
(written as a normal three vector) is, again by help of (\ref{eq:15}),
derived in the concise form:
\begin{equation}
\label{eq:21}
 \mathbf{F}_{\mu\nu} = \partial_\mu \mathbf{A}_\nu - \partial_\nu \mathbf{A}_\mu
 + 2g \mathbf{A}_\mu \times \mathbf{A}_\nu.
\end{equation}

As is well known from non-Abelian gauge theory \cite{yang,lee}, the
non-commuting nature of the vector components of the isospin $\mathbf{S}$
yields the nonlinear terms in the field tensor (\ref{eq:21}). The Lagrangian
of the present SU(2)-like gauge field (see text books) theory is defined by
\begin{equation}
\label{eq:20}
\mathcal{L}_\mathrm{F} = - \frac{1}{16} \mathrm{Tr}( F_{\mu\nu} F^{\mu\nu} )
= - \frac{1}{4} \, (\mathbf{F}_{\mu\nu} \cdot \mathbf{F}^{\mu\nu}).
\end{equation}
The term on the right of (\ref{eq:20}) comes from the properties of the
isospin matrices in (\ref{eq:16}), which also obey the relation:
$\mathrm{Tr}(\bar{S}_i\bar{S}_j)=-4\delta_{i,j}$. This produces the
normalization factor 16 in $\mathcal{L}_\mathrm{F}$ instead of 2 of the
standard SU(2) gauge group in its fundamental representation.

The two combined equations (\ref{eq:18}) and (\ref{eq:20}) define the total
Lagrangian, $\mathcal{L} = \mathcal{L}_\mathrm{F} + \mathcal{L}_\mathrm{I} +
\mathcal{L}_\mathrm{M}$, and govern the dynamics of the fermionic Majorana
field and the associated bosonic gauge fields. Variation of the Yang-Mills
[15] action four-integral over $\mathcal{L}$ with respect to the relevant
field variables finally yields the two real coupled partial-differential
matrix equations:

\begin{equation}
\label{eq:24}
(\bar{\gamma}^\mu \partial_\mu + m ) \Psi(x)
= g \bar{\gamma}^\mu  \mathbf{A}_\mu(x) \cdot \bar{\mathbf{S}} \Psi(x),
\end{equation}

\begin{equation}
\label{eq:25}
\partial_\mu \mathbf{F}^{\mu \nu}(x) + 2g (\mathbf{A}_\mu(x) \times \mathbf{F}^{\mu\nu}(x))
= g \Psi^\mathrm{T}(x) \bar{\alpha}^\nu \mathbf{S} \Psi(x).
\end{equation}

Remember that $\bar{\gamma}^\mu=\bar{\beta}\bar{\alpha}^\mu$. Both equations
(\ref{eq:24}) and (\ref{eq:25}) have terms linear in the fields (free fields)
and quadratic ones describing the field coupling. The gauge field equation
(\ref{eq:25}) in addition contains a term quadratic in this field, a property
leading to nonlinear interaction of the field with itself.

In the next section we will make some drastic approximations to the above
complex system of coupled field equations, yet which provide an interesting
simplified model which already seems to indicate the possibility of flavour
mixing and oscillations at frequencies determined by the effective masses as
induced by mean-gauge-field coupling.

\section{Flavour mixing and neutrino oscillations}
\label{s5}

Let us consider a very much simplified version of the general gauge-field
model equations (\ref{eq:24}) and (\ref{eq:25}). Assume a fixed location and
time dependence only, and further consider a mean-field approach to the gauge
fields, thereby assuming a static field, with the vector
$\mathbf{A}_\mu(x)=(\mathbf{A},0,0,0)$ being constant. Then $\bar{\gamma}^\mu
\mathbf{A}_\mu= \bar{\beta}\mathbf{A}$. Dimensionally, the coupling constant
$g$ times the field $\mathbf{A}$ must correspond to a mass, and thus we may
write $g \mathbf{A}\cdot\mathbf{S}$ as a field-related mass matrix, which by
help of (\ref{eq:16}) takes the form:
\begin{equation}
\label{eq:120}
\bar{M}=
\left(
\begin{array}{cccc}
0 & -M_\mathrm{y} & -M_\mathrm{z} & -M_\mathrm{x} \\
+M_\mathrm{y} & 0 & -M_\mathrm{x} & +M_\mathrm{z} \\
+M_\mathrm{z} & +M_\mathrm{x} & 0 & -M_\mathrm{y} \\
+M_\mathrm{x} & -M_\mathrm{z} & M_\mathrm{y}  & 0 \\
\end{array}
\right),
\end{equation}
which obeys $\bar{M}^\mathrm{T}=-\bar{M}$, where the superscript $\mathrm{T}$
indicates the transposed matrix, and which reflects the properties of the
real Majorana field spin operator. Then the spinor time evolution is given by
the mixing equation:
\begin{equation}
\label{eq:121}
\left( \bar{\beta}(\frac{\partial}{\partial t} + \bar{M}) + m \right) \Psi(t) = 0.
\end{equation}
In this simple mean-field model, there are four masses corresponding to the
basic mass $m$ and three others (inertia induced by mean-gauge-field
coupling). The spinor solution involves oscillations or flavour mixing at the
frequencies to be determined from the solution of the eigenvalue problem
posed by equation (\ref{eq:121}). We can rewrite it as a standard
second-order oscillation equation as follows. Multiplying the equations, i.e.
their wavefunctions, by $\bar{\beta}$ yields two coupled equations for $\Psi$
and $\bar{\beta}\Psi$, which can be inserted into each other to obtain a
coupled oscillator (remember there are four flavour degrees of freedom)
equation as follows:
\begin{equation}
\label{eq:122}
\left( \frac{\partial^2}{\partial t^2} + 2 \bar{M} \frac{\partial}{\partial t}
-  (M_\mathrm{x}^2 + M_\mathrm{y}^2 + M_\mathrm{z}^2) + m^2 \right) \Psi(t) = 0.
\end{equation}
Without the field coupling we have independent harmonic oscillations with a
phase $mt$, with it we obtain mixing, which is associated with frequency
splitting and damping or excitation in dependence upon the mean static
gauge-field vector $\mathbf{M}=(M_\mathrm{x},M_\mathrm{y},M_\mathrm{z})$.

We may think that the mass term could be replaced with a diagonal mass matrix
representing different intrinsic masses of the four flavours, i.e. we may
define
\begin{equation}
\label{eq:123}
\bar{m} =
\left(
\begin{array}{cccc}
m_1 & 0 & 0 & 0 \\
0 & m_2 & 0 & 0 \\
0 & 0 & m_3 & 0 \\
0 & 0 & 0 & m_4\\
\end{array}
\right).
\end{equation}
However, such a matrix $\bar{m}$ does not commute with $\bar{M}$, and the
SU(2) flavour gauge symmetry would be broken by adopting it. Therefore, we
will not consider this case any further for the sake of simplicity. But we
consider now spatial variations, given again a constant background gauge
field, and add the spatial derivative term to describe spatial propagation,
and thus obtain:
\begin{equation}
\label{eq:124} \left( \frac{\partial}{\partial t} + \bar{
\boldsymbol{\alpha}} \cdot \frac{\partial}{\partial \mathbf{x}} \right)
\Psi(\mathbf{x},t) = (\bar{\beta} m + \bar{M}) \Psi(\mathbf{x},t),
\end{equation}
which differs from the free Majorana field (\ref{eq:6}) essentially by the
gauge-field-induced mass mixing term. Considering again $\Psi$ and
$\bar{\beta}\Psi$, and by eliminating one them, we obtain
\begin{equation}
\label{eq:125}
\left( \frac{\partial^2}{\partial t^2} - \frac{\partial^2}{\partial \mathbf{x}^2} + 2 \bar{M} \frac{\partial}{\partial t}
+ \bar{M}^2 + m^2 \right) \Psi(\mathbf{x},t) = 0.
\end{equation}
Use has been made of the property $(\bar{\boldsymbol{\alpha}} \cdot
\mathbf{D})^2= \mathbf{D}^2$ for any vector $\mathbf{D}$, and that the alphas
and beta anticommute. Since the equation is real, we can assume a harmonic
wave superposition and make for the solution the ansatz:
\begin{equation}
\label{eq:126}
\Psi(\mathbf{x},t) = \mathfrak{C} \cos(Et - \mathbf{p}\cdot\mathbf{x})
+ \mathfrak{S} \sin(Et - \mathbf{p}\cdot\mathbf{x}),
\end{equation}
with four-component amplitude or flavour-polarisation spinors $\mathfrak{C}$
and $\mathfrak{S}$. Insertion of this ansatz yields the coupled set of
equations:
\begin{equation}
\label{eq:127}
(-E^2+\mathbf{p}^2+m^2-M^2) \mathfrak{C} - 2E\bar{M} \mathfrak{S}=0,
\end{equation}
\begin{equation}
\label{eq:128}
(-E^2+\mathbf{p}^2+m^2-M^2) \mathfrak{S} + 2E\bar{M} \mathfrak{C}=0,
\end{equation}
where $M=\sqrt{-1/4\mathrm{Tr}(\bar{M}^2)}=\sqrt{\mathbf{M}^2}$. For
nontrivial solutions to exist the determinant of the system (\ref{eq:127})
and (\ref{eq:128}) must vanish, which yields for the energy eigenvalues the
four roots:
\begin{equation}
\label{eq:129}
E_{1,2}= \pm M + \sqrt{\mathbf{p}^2+m^2},
\end{equation}
\begin{equation}
\label{eq:130}
E_{3,4}= \pm M - \sqrt{\mathbf{p}^2+m^2}.
\end{equation}
For zero gauge field one retains the normal relativistic energies of free
particles. Apparently, the constant gauge field yields energy splitting and
flavour mixing, as the new eigenstates are obtained by mixing of the
free-particle eigenstates. The mixed state is determined by the flavour
spinors $\mathfrak{C}$ and $\mathfrak{S}$, which we shall not calculate in
detail. Owing to the application of the gauge field the four flavours
acquired different energies in dependence on the field strength $M$, and thus
the degeneracy of the original (just one mass $m$) energy spectrum has been
lifted.

Finally, we recall that for free Majorana particles of each flavour species
the polarization spinors depend on energy and momentum, according to the
eigenfunctions given in appendix Section~\ref{s7}. For the free Majorana
field, when being written like in (\ref{eq:126}) in terms of sine and cosine
functions, the corresponding four-component spinors $C$ and $S$ are
determined by the polarization spinor $\tilde{\alpha}$, i.e. $C =
1/\sqrt{2}\,\mathrm{Re}\tilde{\alpha}$ and $S =
1/\sqrt{2}\,\mathrm{Im}\tilde{\alpha}$ for particles, and similarly for
antiparticles. For them we just have to replace $\tilde{\alpha}$ by
$\tilde{\beta}$, whose dependence on energy is given by the coefficient
$\varepsilon_\pm$ as defined in the appendix Section~\ref{s7}. The
corresponding flavour polarization spinors are more complicated, though, and
have in principle to be determined by finding for each energy the
eigenvectors of the full sixteen-dimensional system given by equations
(\ref{eq:127}) and (\ref{eq:128}). This tedious calculation shall not be done
here.

\section{Discussion and conclusion}
\label{s6}

A gauge theory for massive neutral particles has been developed on the basis
of the real four-component Majorana equation. The novel aspect is that by use
of its spin operator a purely imaginary representation of the SU(2) algebra
can be defined, which provides a covariant derivative or connection to the
gauge field that is real. Such a minimal coupling can preserve the real
nature of the Majorana equation. The associated isospin is four-dimensional,
and thus via this procedure we introduce four intrinsic degrees of freedom to
the Majorana field. What could be the nature of these degrees of freedom?

The empirical fact that neutrinos oscillate and thereby change flavour has
motivated us to make such a proposal. The main aim was to describe the
mathematical feasibility of coupling the real Majorana field with a gauge
field, a possibility that was not obvious and, to our best knowledge of the
literature, not known before. Adding a fourth member to the flavour family,
is here just a matter of mathematical necessity and imposed by the choice of
our isospin, but it remains physically speculative. However, flavour could in
this way become a dynamic trait of the neutral Majorana particles, and thus
lead to their linkage and a deeper understanding of neutrino mixing.

We know the standard-model fermions come in three flavours, yet the present
model would imply another fourth flavour, beyond the electron, muon and tau,
and their neutrinos, and perhaps similarly for the quarks as well. The four
flavours result here from the dimension of the representation of the gauge
group, and as such are a cogent consequence if we accept that present
description at all. Its dimension is that of space-time in the covariant
Dirac and Majorana equations, and thus given just by the dimension of the
gamma matrices. The particle quantum states in these equations are ordered
according to past and future (antiparticles and particles) and right-helical
or left-helical. The isospin, when being based on the spin operator that
normally describes rotations in three-dimensional space, may correspond to
the remaining four spatial orientations, which is forward and backward and up
and down. This may be considered a simple intuitive interpretation of the
physical content of the 16-component spinor $\Psi$, which arranges each
flavour family into a quadruplett of real Majorana spinors for neutrinos, or
perhaps complex Dirac spinors for charged leptons or quarks. So the sixteen
components correspond to the sixteen ($2^4$) signed domains of space-time,
and thus the number of flavours must be four but cannot be higher.

The simple static mean-field approximation of the gauge field can produce
Majorana-field flavour oscillations and mixing, can lift the mass degeneracy
of the four Majorana neutrinos and cause a splitting of their mass spectrum,
the size of which is determined by the gauge field strength. So mass is
partly acquired here from gauge-field energy. This simple model just
illustrates the physical potential of a flavour-mixing gauge theory of
massive neutrinos.

\section{Appendix I: Eigenfunctions of the free Majorana equation}
\label{s7}

Let us discuss briefly the eigenfunctions of the free Majorana field, which
obeys equation (\ref{eq:3}) or (\ref{eq:5}) and can be decomposed into its
particle and antiparticle components, and according to the recent papers of
Marsch \cite{marsch1, marsch2} be written
\begin{equation}
\label{eq:101} \psi(\mathbf{x},t) = \psi_\mathrm{P}(\mathbf{x},t) + \psi_\mathrm{A}(\mathbf{x},t).
\end{equation}
These two contributions can be expressed, in terms of the complex
four-component polarization spinors to be defined below, separately as
follows:
\begin{eqnarray}
\label{eq:102}
\psi_\mathrm{P}(\mathbf{x},t) = \exp(-\mathrm{i} E t + \mathrm{i} \mathbf{p}\cdot\mathbf{x})
a(\mathbf{p}) \tilde{\alpha}(\mathbf{p}) + \exp(\mathrm{i} E t - \mathrm{i} \mathbf{p}\cdot\mathbf{x})
a^*(\mathbf{p}) \tilde{\alpha}^*(\mathbf{p}),
\end{eqnarray}
\begin{eqnarray}
\label{eq:103}
\psi_\mathrm{A}(\mathbf{x},t) = \exp(-\mathrm{i} E t + \mathrm{i} \mathbf{p}\cdot\mathbf{x})
b(\mathbf{p}) \tilde{\beta}(\mathbf{p}) + \exp(\mathrm{i} E t - \mathrm{i} \mathbf{p}\cdot\mathbf{x})
b^*(\mathbf{p}) \tilde{\beta}^*(\mathbf{p}).
\end{eqnarray}
Apparently, $\psi_\mathrm{P,A}=\psi^*_\mathrm{P,A}$, and thus the wave
functions are real. The polarization spinors $\tilde{\alpha}$ and
$\tilde{\beta}$ are the two eigenfunctions of the helicity operator
$\boldsymbol{\Sigma}\cdot\hat{\mathbf{p}}$ (as defined in Section~\ref{s3})
with eigenvalues $+1$ and $-1$.
\begin{equation}
\label{eq:110} ({\boldsymbol{\Sigma}} \cdot \hat{\mathbf{p}})
\tilde{\alpha}(\mathbf{p})= + \tilde{\alpha}(\mathbf{p}).
\end{equation}
\begin{equation}
\label{eq:111} ({\boldsymbol{\Sigma}} \cdot \hat{\mathbf{p}})
\tilde{\beta}(\mathbf{p})= - \tilde{\beta}(\mathbf{p}).
\end{equation}
After quantization the complex Fourier amplitudes $a^*(\mathbf{p})$ and
$b^*(\mathbf{p})$ turn into creation operators of particles and
antiparticles, which on average (in the sense of a quantum-field expectation
value the Majorana field) have opposite helicities. For the sake of
completeness we give here the full four-component polarization spinors, which
in terms of the angles of the momentum unit vector $\hat{\mathbf{p}}=
(\sin\theta \cos\phi, \sin\theta \sin\phi, \cos\theta)$ and of the module $p$
read as follows:
\begin{equation}
\label{eq:105}
\tilde{\alpha}(\mathbf{p}) = \frac{1}{2} \left(
\begin{array}{c}
(+\varepsilon_+(p)  \cos \frac{\theta}{2} \; \mathrm{e}^{-\frac{\mathrm{i}}{2} \phi} + \varepsilon_-(p) \sin \frac{\theta}{2} \; \mathrm{e}^{+\frac{\mathrm{i}}{2} \phi})(1-\mathrm{i}) \\
(+\varepsilon_+(p)  \sin \frac{\theta}{2} \; \mathrm{e}^{+\frac{\mathrm{i}}{2} \phi} - \varepsilon_-(p) \cos \frac{\theta}{2} \; \mathrm{e}^{-\frac{\mathrm{i}}{2} \phi})(1-\mathrm{i}) \\
(+\varepsilon_+(p)  \cos \frac{\theta}{2} \; \mathrm{e}^{-\frac{\mathrm{i}}{2} \phi} - \varepsilon_-(p) \sin \frac{\theta}{2} \; \mathrm{e}^{+\frac{\mathrm{i}}{2} \phi})(1+\mathrm{i}) \\
(+\varepsilon_+(p)  \sin \frac{\theta}{2} \; \mathrm{e}^{+\frac{\mathrm{i}}{2} \phi} + \varepsilon_-(p) \cos \frac{\theta}{2} \; \mathrm{e}^{-\frac{\mathrm{i}}{2} \phi})(1+\mathrm{i}) \\
\end{array}
\right).
\end{equation}
\begin{equation}
\label{eq:106}
\tilde{\beta}(\mathbf{p}) = \frac{1}{2} \left(
\begin{array}{c}
(-\varepsilon_-(p) \sin \frac{\theta}{2} \; \mathrm{e}^{-\frac{\mathrm{i}}{2} \phi} + \varepsilon_+(p) \cos \frac{\theta}{2} \; \mathrm{e}^{+\frac{\mathrm{i}}{2} \phi})(1-\mathrm{i}) \\
(+\varepsilon_-(p)  \cos \frac{\theta}{2} \; \mathrm{e}^{+\frac{\mathrm{i}}{2} \phi} + \varepsilon_+(p) \sin\frac{\theta}{2} \; \mathrm{e}^{-\frac{\mathrm{i}}{2} \phi})(1-\mathrm{i}) \\
(-\varepsilon_-(p) \sin \frac{\theta}{2} \; \mathrm{e}^{-\frac{\mathrm{i}}{2} \phi} - \varepsilon_+(p) \cos \frac{\theta}{2} \; \mathrm{e}^{+\frac{\mathrm{i}}{2} \phi})(1+\mathrm{i}) \\
(+\varepsilon_-(p)  \cos \frac{\theta}{2} \; \mathrm{e}^{+\frac{\mathrm{i}}{2} \phi} - \varepsilon_+(p) \sin \frac{\theta}{2} \; \mathrm{e}^{-\frac{\mathrm{i}}{2} \phi})(1+\mathrm{i}) \\
\end{array}
\right).
\end{equation}
It was convenient to introduce the two real quantities
\begin{equation}
\label{eq:107}
\varepsilon_\pm (p)= \sqrt{\frac{E(p) \pm p}{2E(p)} },
\end{equation}
the squares of which add up to unity, $\varepsilon_1^2 + \varepsilon_2^2 =1$.
The relativistic energy of a particle is $E(p) = \sqrt{m^2+p^2}$. A useful
property of the epsilons is the obvious relation $(E \pm p)\varepsilon_\mp =
m \varepsilon_\pm$. Using their properties, one can readily show that
$\tilde{\alpha}^\dag \tilde{\alpha}=1$ and $\tilde{\beta}^\dag
\tilde{\beta}=1$, respectively, $\tilde{\alpha}^\dag \tilde{\beta}=0$ and
$\tilde{\beta}^\dag \tilde{\alpha}=0$. If $m=0$, then $\varepsilon_+=1$ and
$\varepsilon_-=0$. It should be emphasized that the above spinors
$\psi_\mathrm{A}$ and $\psi_\mathrm{P}$ always are a superposition of both
helicity states, as it is required for a massive relativistic particle. The
eigenvalue equation of the helicity operator (based on the spin operator, see
Section~\ref{s3}) in Fourier space reads for the complex conjugated
polarization spinor:
\begin{equation}
\label{eq:108} ({\boldsymbol{\Sigma}} \cdot \hat{\mathbf{p}})
\tilde{\alpha}^*(\mathbf{p})= - \tilde{\alpha}^*(\mathbf{p}).
\end{equation}
Similarly, for $\tilde{\beta}^*(\mathbf{p})$, which has the opposite positive
helicity, we obtain:
\begin{equation}
\label{eq:109} ({\boldsymbol{\Sigma}} \cdot \hat{\mathbf{p}})
\tilde{\beta}^*(\mathbf{p})= + \tilde{\beta}^*(\mathbf{p}).
\end{equation}
By multiplying out explicitly all possible scalar products involving the two
complex conjugated polarization vectors $\tilde{\alpha}^*(\mathbf{p})$ and
$\tilde{\beta}^*(\mathbf{p})$, one finds further that $\tilde{\alpha}^\dag
\tilde{\alpha}^*=(\tilde{\alpha}^\mathrm{T} \tilde{\alpha})^*=0$ and
$\tilde{\beta}^\dag \tilde{\beta}^*=(\tilde{\beta}^\mathrm{T}
\tilde{\beta})^*=0$, but $\tilde{\alpha}^\dag
\tilde{\beta}^*=(\tilde{\alpha}^\mathrm{T}
\tilde{\beta})^*=\mathrm{i}(\varepsilon_+^2-\varepsilon_-^2)$, and similarly
$\tilde{\beta}^\dag \tilde{\alpha}^*=(\tilde{\beta}^\mathrm{T}
\tilde{\alpha})^*=\mathrm{i}(\varepsilon_+^2-\varepsilon_-^2)$. Consequently,
there are four independent real polarization vectors spanning the full
Hilbert space of the Majorana equation. The eigenvalues $\pm E(p)$ are
two-fold degenerate, and their subspaces are spanned by
$\tilde{\alpha}(\mathbf{p})$ and $\tilde{\alpha}^*(\mathbf{p})$, respectively
$\tilde{\beta}(\mathbf{p})$ and $\tilde{\beta}^*(\mathbf{p})$. Both
helicities are needed because of the finite mass of the relativistic Majorana
particles. However, the degeneracy concerning the replacement of the spin
$\boldsymbol{\Sigma}$ by its negative (spin reversal) has been lifted,
because the real Majorana equation does not obey chirality conjugation (see
again Section~\ref{s3}).

\section{Appendix II: Two Majorana equations}
\label{s8}

As was mentioned already in the main text, the two-component complex
(\ref{eq:5}) and four-component real (\ref{eq:6}) Majorana equations are
fully equivalent. However, as discussed by Marsch \cite{marsch1, marsch2},
there is a second chirality-conjugated equation obtained by operating with
$\tau$ on (\ref{eq:5}), which yields
\begin{equation}
\label{eq:300} \mathrm{i}\left( \frac{\partial}{\partial t} -
\boldsymbol{\sigma} \cdot \frac{\partial}{\partial \mathbf{x}} \right)
\chi(\mathbf{x},t) = -m \tau \chi(\mathbf{x},t),
\end{equation}
with $\chi = \tau \phi$, which is the left-chiral counterpart to the
right-chiral field $\phi$. This equation can be obtained by simply inverting
the sign of the Pauli matrices in (\ref{eq:5}), including of course of $\tau$
as it contains $\sigma_\mathrm{y}$. The $\pm$ signs in front of the Pauli
matrices reflect the reducibility of the Lorentz group, which can in fact be
decomposed into its left- and right-chiral components. In the chiral or Weyl
representation the charge conjugation operator is given by
$\mathcal{C}=\gamma_\mathrm{y} \mathbb{C}= \gamma\tau$, which when being
squared equals unity. So its eigenvalues are $\pm$, and its eigenfunctions
obey $\mathcal{C}\psi^\mathcal{C}_\pm=\pm\psi^\mathcal{C}_\pm$. The
eigenfunction with positive eigenvalue obeys the Dirac equation and in Weyl
representation is given by the spinor:
\begin{equation}
\label{eq:301}
\psi^\mathcal{C}
=\left(
\begin{array}{c}
\phi \\
\tau \phi\\
\end{array}
\right)=
\left(
\begin{array}{c}
\phi \\
\chi\\
\end{array}
\right).
\end{equation}
Insertion of this spinor yields for the two two-component spinors $\phi$ and
$\chi$ the twin Majorana equations (\ref{eq:5}) and (\ref{eq:300}), which are
coupled through the restricting condition $\chi=\tau\phi$, which according to
(\ref{eq:301}) guarantees chirality self-conjugation.

Correspondingly, we obtain a second real four-component Majorana equation by
simply taking the negatives of the matrices $\alpha, \beta, \gamma$ of
equation (\ref{eq:3}). This choice transposes the vector
$\bar{\boldsymbol{\alpha}}$ of (\ref{eq:7}) to its negative and also gives a
negative $\bar{\beta}$, so that we obtain the second real Majorana equation
as
\begin{equation}
\label{eq:302} \left( \frac{\partial}{\partial t} - \bar{
\boldsymbol{\alpha}} \cdot \frac{\partial}{\partial \mathbf{x}} \right)
\tilde{\psi}(\mathbf{x},t) = -m \bar{\beta}\, \tilde{\psi}(\mathbf{x},t).
\end{equation}
The solution is according to the Table~\ref{table:1} given by the chirally
conjugated spinor
$\tilde{\psi}(\mathbf{x},t)=\bar{\delta}\psi(\mathbf{x},t)$, where $\psi$
solves the Majorana equation (\ref{eq:6}). Note that both Majorana equations
obey the parity and time-reversal symmetry, but individually break by
construction chirality conjugation which links them together. Concerning the
spin operator (\ref{eq:9}), it will also be transposed for the second
conjugated Majorana field and turn into the negative of the spin of the
primary field, i.e. $\tilde{\mathbf{S}}= - \mathbf{S}$. However, the
connection to the gauge field after (\ref{eq:14}) will not change, as once we
have made our choice of the symmetry group their generators are given and do
not undergo the discussed space-time symmetry operations.

Returning to the twin two-component Majorana equations (\ref{eq:5}) and
(\ref{eq:300}), we recall that in the chiral Weyl representation of the Dirac
equation the right- and left-chiral field are obtained by projection with the
help of the chirality operator $\gamma^5$. Correspondingly, the Dirac field
can be decomposed into its right- (index R) and left-chiral (index L)
components, such that
\begin{equation}
\label{eq:304}
\psi_\mathrm{R}
=\frac{1}{\sqrt{2}}\left(
\begin{array}{c}
\phi_\mathrm{R} \\
0 \\
\end{array}
\right)
\;\;\;\; \mathrm{and} \;\;\;\;
\psi_\mathrm{L}
=\frac{1}{\sqrt{2}} \left(
\begin{array}{c}
0 \\
\phi_\mathrm{L}\\
\end{array}
\right),
\end{equation}
where we introduced the front factor for normalization, assuming that the
functions $\psi_\mathrm{R}$ and $\psi_\mathrm{L}$ are individually normalized
to unity. Above we used the symbols $\phi_\mathrm{R}=\phi$ and
$\phi_\mathrm{L}=\chi$, with the chirality self-conjugation constraint that
$\chi=\tau\phi$ to be kept in mind. Therefore the mass term mixing the two
chiral components, such that in the Lagrange density we have $m
\bar{\psi}\psi=m/2(\phi^\dag\chi + \chi^\dag\phi)$, can formally be made
diagonal by exploiting the above condition. The price to be payed for this
decoupling is that the mass term becomes instead of trivial multiplication a
nontrivial operator involving with $\tau$ also the inconvenient
complex-conjugation operation $\mathbb{C}$. As the result we get the Lagrange
density
\begin{equation}
\label{eq:305} \mathcal{L} = \frac{1}{2}\left(\phi^\dag_\mathrm{R}
\left( \mathrm{i}\sigma_\mathrm{R}^\mu \partial_\mu - m \tau_\mathrm{R}\right)
\phi_\mathrm{R} + \phi^\dag_\mathrm{L}
\left( \mathrm{i}\sigma_\mathrm{L}^\mu \partial_\mu - m \tau_\mathrm{L}\right)
\phi_\mathrm{L}\right).
\end{equation}
It can be shown (see the paper by Pal \cite{pal}) to be hermitian, while the
two terms being their hermitian conjugates, which can be validated by using
the relations $\phi_\mathrm{R}=-\tau\phi_\mathrm{L}$, respectively
$\phi_\mathrm{L}=\tau\phi_\mathrm{R}$, which are constitutive for the complex
two-component Majorana field. Here we defined for the sake of formal symmetry
the symbols: $\tau_\mathrm{R}=\tau$, $\tau_\mathrm{L}=-\tau$, with $\tau=
\sigma_\mathrm{y} \mathbb{C}$, and $\sigma^\mu_\mathrm{R}=(1,
\boldsymbol{\sigma})$ and $\sigma^\mu_\mathrm{L}=(1, -\boldsymbol{\sigma})$.
However, with $\sigma^\mu=\sigma^\mu_\mathrm{R}$ we may also rewrite
(\ref{eq:305}) in terms of $\phi=\phi_\mathrm{R}$ as follows
\begin{equation}
\label{eq:308} \mathcal{L} = \mathrm{i} \, \mathrm{Im}\left(\phi^\dag
\left( \mathrm{i}\sigma^\mu \partial_\mu - m \tau\right)
\phi \right),
\end{equation}
which emphasizes that there is only a single two-component complex Majorana
field.

The result of equation (\ref{eq:305}) can now be directly transferred to the
real Majorana equation, following the same mathematics that lead to equation
(\ref{eq:6}), respectively (\ref{eq:302}). By decomposing the two parts of
(\ref{eq:305}) into their real and imaginary parts and writing the density in
terms of four-component real spinors after \cite{marsch2}, the resulting
alpha and beta matrices will of course change correspondingly, as we obtained
$\bar{\boldsymbol{\alpha}}_\mathrm{R} = \bar{\boldsymbol{\alpha}}$ and
$\bar{\boldsymbol{\alpha}}_\mathrm{L} = -\bar{\boldsymbol{\alpha}}$,
respectively, $\bar{\beta}_\mathrm{R} = \bar{\beta}$ and
$\bar{\beta}_\mathrm{L}= -\bar{\beta}$. By definition, we obtain
$\bar{\alpha}^\mu_\mathrm{R}=(1, \bar{\boldsymbol{\alpha}})$ and
$\bar{\alpha}^\mu_\mathrm{L}=(1, -\bar{\boldsymbol{\alpha}})$. Consequently,
$\bar{\alpha}^\mu_\mathrm{R}=\bar{\alpha}_{\mathrm{L}\,\mu}$, which shows how
chirality and space-time are intrinsically connected. We can then write the
combined Majorana Lagrange density as
\begin{equation}
\label{eq:306}
\mathcal{L} = \frac{\mathrm{i}}{2}\left(\Psi_\mathrm{R}^\mathrm{T}(\bar{\alpha}_\mathrm{R}^\mu \partial_\mu
- \bar{\beta}_\mathrm{R} m )\Psi_\mathrm{R}
+ \Psi_\mathrm{L}^\mathrm{T}(\bar{\alpha}_\mathrm{L}^\mu \partial_\mu
 - \bar{\beta}_\mathrm{L} m )\Psi_\mathrm{L}\right).
\end{equation}
This equation shows a formal symmetry between the left-chiral and
right-chiral components of the Majorana fields which at first sight seem to
be independent. However, they are not, as we have $\psi_\mathrm{L}=
\bar{\delta} \psi_\mathrm{R}$, and vice versa $\psi_\mathrm{R}= -\bar{\delta}
\psi_\mathrm{L}$. Also recall that $\{\bar{\delta},\bar{\beta}\}=0$ and
$\{\bar{\delta},\bar{\boldsymbol{\alpha}}\}=\mathbf{0}$. Note that by its
definition $\delta^\mathrm{T}=-\delta$ and $\delta^2=-1$. So by inserting
$\psi_\mathrm{L}= \bar{\delta} \psi_\mathrm{R}$ into the above Lagrangian,
the second term turns out to be exactly equal to the first. So there is only
a single real Majorana field the Lagrangian of which was already given in
(\ref{eq:9}). Therefore, considering the real Majorana field involving real
four-component spinors, we believe, is advantageous over the complex Majorana
field involving two-component complex spinors and requires to take care of
the mathematical subtlety and complication of the operator $\tau$.

\section{Appendix III: Chiral decomposition of the spin operator}
\label{s9}

Concerning the possible choice of the gauge symmetry group, traditionally
SU(N) is used with the prominent SU(2) and SU(3) Lie groups employed in their
fundamental representations for the weak and strong interactions. However,
the real SU(2) representation chosen in (\ref{eq:16}) is special in so far as
it corresponds to the physical quantity angular momentum of the Majorana
field. So there is no need to use left- or right-chiral Weyl fields, if the
neutrino is assumed to be a massive Majorana particle, which by its very
nature comes as a left-helical particle and right-helical antiparticle. The
Majorana spin operator $\mathbf{S}_\mathrm{M}$ was adopted as isospin in the
present work. It can be written in terms of the Pauli matrices in the form:
\begin{equation}
\label{eq:16a}
\mathbf{S}_\mathrm{M} = \frac{1}{2} \left(
\left(
\begin{array}{cc}
0 & -\mathrm{i} \sigma_\mathrm{x}\\
\mathrm{i}\sigma_\mathrm{x} &  0 \\
\end{array}
\right),
\left(
\begin{array}{cc}
\sigma_\mathrm{y} & 0 \\
0 & \sigma_\mathrm{y} \\
\end{array}
\right),
\left(
\begin{array}{cc}
0 & -\mathrm{i} \sigma_\mathrm{z}\\
\mathrm{i} \sigma_\mathrm{z} &  0 \\
\end{array}
\right) \right).
\end{equation}

We like to discuss then how to couple Dirac fermions (if also carrying the
charge $g$) to the same gauge field as discussed in Section~\ref{s3}. The
procedure how to do this is not at all obvious. Like in the weak
interactions, perhaps a projection onto chiral eigenfunctions is needed for
Dirac fermions. This is rather speculative and needs further investigations.
As is well known the spin operator of the Dirac equation is given, both in
Dirac and Weyl representation, by the expression:
\begin{equation}
\label{eq:28}
\mathbf{S}_\mathrm{D} = \frac{1}{2} \left(
\left(
\begin{array}{cc}
\sigma_\mathrm{x} & 0 \\
0 & \sigma_\mathrm{x} \\
\end{array}
\right),
\left(
\begin{array}{cc}
\sigma_\mathrm{y} & 0 \\
0 & \sigma_\mathrm{y} \\
\end{array}
\right),
\left(
\begin{array}{cc}
\sigma_\mathrm{z} & 0 \\
0 & \sigma_\mathrm{z} \\
\end{array}
\right) \right).
\end{equation}
which is reducible and fully described by the fundamental SU(2) group as
defined by the Pauli matrices. Here we want to mention that the spin can be
decomposed by means of the chiral operator $\gamma^5$ and its related
projection operators $P_{\mathrm{R},\mathrm{L}}= 1/2(1 \pm \gamma^5)$. They
are idempotent, i.e. $P_{\mathrm{R},\mathrm{L}}^2=P_{\mathrm{R},\mathrm{L}}$,
provide a decomposition of unity: $P_\mathrm{R}+P_\mathrm{L}=1$, and are
orthogonal by construction: $P_\mathrm{R} P_\mathrm{L}=0$. Furthermore,
$\gamma^5$ commutes with the spin operator, which is derived from the
commutator of two gamma matrices after (\ref{eq:10}). In fact it commutes
with $\mathbf{S}$ in any representation. Therefore, we can write for the spin
operator $\mathbf{S}$, which obeys
$\mathbf{S}\times\mathbf{S}=\mathrm{i}\mathbf{S}$, formally
\begin{equation}
\label{eq:307}
\mathbf{S}= (P_\mathrm{R} + P_\mathrm{L})\mathbf{\mathbf{S}}=
\mathbf{S}_\mathrm{R} + \mathbf{S}_\mathrm{L},
\end{equation}
and consequently obtain
$\mathbf{S}_{\mathrm{R},\mathrm{L}}\times\mathbf{S}_{\mathrm{R},\mathrm{L}}=\mathrm{i}\mathbf{S}_{\mathrm{R},\mathrm{L}}$,
and $[ \mathbf{S}_\mathrm{R}, \mathbf{S}_\mathrm{L} ]=0$. For the Majorana
field the spin operator (\ref{eq:11}) is purely imaginary and the projected
components become complex, which is not helpful. In the chiral Weyl
representation we simply get
$\mathbf{S}_{\mathrm{R},\mathrm{L}}=P_{\mathrm{R},\mathrm{L}}\boldsymbol{\sigma}/2$,
which is irreducible and simplest. So the question comes up which the
adequate representation is to be used for the SU(2) symmetry group.

If we assume that any Dirac fermion is like a Majorana fermion endowed with
four internal flavour degrees of freedom, then its interaction is mediated by
the same gauge field $\mathbf{A}_\mu$ considered for the Majorana fermion. We
may take the isospin to be given consistently by the $\mathbf{S}$ of
(\ref{eq:16}), and the coupling to the field be given by the same connection
(\ref{eq:14}). However, we may prefer $\mathbf{S}=\mathbf{S}_\mathrm{D}$ to
be used in the covariant derivative, i.e. favour the connection
\begin{equation}
\label{eq:29}
D_\mu  = \partial_\mu + \mathrm{i} g \mathbf{S}_\mathrm{D} \cdot \mathbf{A}_\mu,
\end{equation}
in the Dirac equation. Formally, such a gauge-field interaction model seems
to resemble the weak interaction theory \cite{weinberg, glashow, salam} of
the standard model, yet there are three major differences. First, when
accepting the connection (\ref{eq:29}) for Dirac fermions, there is no
restriction of the gauge-field-coupling to the left-chiral fermion field
components only. Such a constraint need not be put anyway on a massive
Majorona field, since it is chirally irreducible and by definition reduced to
two degrees of freedom in comparison with a Dirac fermion. Secondly, both
fermion species can be massive (with a single mass for all flavours),
implying though that chiral symmetry is not obeyed. Thirdly, by construction
of such a model, it will lead to lepton mixing and gauge-field mediated
interactions among all fermions with different flavour, which are here
assembled into unconstrained quadruplets, but not into doublets or singlets
defined by chiral projection.

Therefore, concerning possible choices of the isospin operator, we could make
use of the individual spins as derived from the genuine but different gamma
matrices in the Majorana and Dirac representations. Correspondingly, the
covariant derivative would employ the respective spin operator of the field
considered, which is then used as isospin providing the coupling to the
common gauge field. This appears to be mathematically feasible, yet the
physical implications remain unclear. To discuss these issues in depth is
beyond the scope and intention of the present paper.

\newpage


\end{document}